\documentclass[journal]{IEEEtran}
\usepackage[table]{xcolor}
\usepackage{amsmath,amsfonts}
\usepackage{algorithm}
\usepackage{algorithmicx}
\usepackage{algpseudocode}
\usepackage{array}
\usepackage[caption=false,font=normalsize,labelfont=sf,textfont=sf]{subfig}
\usepackage{textcomp}
\usepackage{stfloats}
\usepackage{url}
\usepackage{xcolor}
\usepackage{verbatim}
\usepackage{graphicx}
\usepackage{cite}
\usepackage{verbatim}
\usepackage{multirow}
\usepackage{makecell}
\usepackage{booktabs}
\usepackage{threeparttable}
\usepackage{amssymb}
\usepackage{bbding}
\usepackage{listings}
\usepackage{float}
\usepackage{changepage}

\begin{document}

\author{Ming~Cheng,
		Yuke~Lin,
        Ming~Li, \IEEEmembership{Senior Member, IEEE}% < stops an unwanted space
\IEEEcompsocitemizethanks{
	\IEEEcompsocthanksitem Ming~Cheng, Yuke~Lin and Ming~Li are with the School of Computer Science, Wuhan University, Wuhan 430072, China, and also with Suzhou Municipal Key Laboratory of Multimodal Intelligent Systems, Digital Innovation Research Center, Duke Kunshan University, Kunshan 215316, China.}% < stops an unwanted space
\thanks{Corresponding author: Ming Li, E-mail: ming.li369@dukekunshan.edu.cn}
}

\title{Sequence-to-Sequence Neural Diarization with Automatic Speaker Detection and Representation}

\maketitle

\begin{abstract}

This paper proposes a novel Sequence-to-Sequence Neural Diarization (S2SND) framework to perform online and offline speaker diarization. It is developed from the sequence-to-sequence architecture of our previous target-speaker voice activity detection system and then evolves into a new diarization paradigm by addressing two critical problems. 1) Speaker Detection: The proposed approach can utilize partially given speaker embeddings to discover the unknown speaker and predict the target voice activities in the audio signal. It does not require a prior diarization system for speaker enrollment in advance. 2) Speaker Representation: The proposed approach can adopt the predicted voice activities as reference information to extract speaker embeddings from the audio signal simultaneously. The representation space of speaker embedding is jointly learned within the whole diarization network without using an extra speaker embedding model. During inference, the S2SND framework can process long audio recordings blockwise. The detection module utilizes the previously obtained speaker-embedding buffer to predict both enrolled and unknown speakers' voice activities for each coming audio block. Next, the speaker-embedding buffer is updated according to the predictions of the representation module. Assuming that up to one new speaker may appear in a small block shift, our model iteratively predicts the results of each block and extracts target embeddings for the subsequent blocks until the signal ends. Finally, the last speaker-embedding buffer can re-score the entire audio, achieving highly accurate diarization performance as an offline system. Experimental results show that our proposed S2SND framework achieves new state-of-the-art diarization error rates (DERs) for online inference on the DIHARD-II (24.41\%) and DIHARD-III (17.12\%) evaluation sets without using oracle voice activity detection. At the same time, it also refreshes the state-of-the-art performance for offline inference on these benchmarks, with DERs of 21.95\% and 15.13\%, respectively.

\end{abstract}

\begin{IEEEkeywords}
Speaker Diarization, Online Speaker Diarization, Sequence-to-Sequence Neural Diarization
\end{IEEEkeywords}

\section{Introduction}

\IEEEPARstart{S}{peaker} diarization aims to split the conversational audio signal into segments with labeled identities, solving the problem of ``Who-Spoke-When''~\cite{park2022review}. It is the core front-end speech processing technique in various downstream tasks like multi-speaker speech recognition, etc~\cite{kanda2020joint}.

Early speaker diarization studies have widely investigated the cascaded methods that process audio signals through a series of independent modules~\cite{6518171,6633085,7078610,wang2018speaker,lin2019lstm,landini2022bayesian,wang2022similarity}. Later, End-to-End Neural Diarization (EEND) methods~\cite{fujita2019end_1,fujita2019end_2, horiguchi2020end,horiguchi2022encoder} are proposed to estimate multiple speakers' voice activities as multi-label classification, where the end-to-end model architecture can be directly optimized by the permutation-invariant training (PIT)~\cite{kolbaek2017multitalker}. Also, Target-Speaker Voice Activity Detection (TSVAD) approaches~\cite{medennikov2020target,wang2022similarity,cheng2023target} combine the advantages of cascaded methods and end-to-end neural networks. A typical TSVAD-based system requires a prior diarization system (e.g., the cascaded method) to extract each speaker's acoustic footprint as the speaker enrollment. Then, a neural network-based module predicts all speakers' corresponding voice activities. This two-stage framework demonstrates promising performance in popular benchmarks such as DIHARD-III~\cite{wang2021ustc} and VoxSRC21-23~\cite{wang2021dku,wang2022dku,cheng2023dku,huh2024vox}.

However, the diarization systems mentioned above are natively designed to process pre-recorded audio offline, which means they cannot satisfy scenarios with low latency demand (e.g., real-time meeting transcription)~\cite{park2022review}. For online speaker diarization, cascaded methods must modify all the built-in components to be capable of online inference, especially the inherent clustering algorithms~\cite{dimitriadis2017developing,zhang22_odyssey}. Online EEND systems are implemented by only replacing the network architecture~\cite{han2021bw} or using a buffer to trace the previous input-output pairs~\cite{xue2021online_1,xue2021online_2,horiguchi2022online}. However, the speaker permutation problem is prone to be affected by the increasing number of speakers in long-form audios, which remains a challenge that has yet to be fully addressed. Although the recent FS-EEND~\cite{liang2024frame} method can determine the speaker permutation according to their appearance order in online scenarios, the error accumulation with inference time may become a new issue. As the post-processing approach, TSVAD models natively process the audio signals blockwise except for acquiring pre-extracted speaker embeddings from the initial stage. Therefore, online TSVAD methods~\cite{wang22j_interspeech, wang2023end} are proposed to enable self-generated speaker embeddings during blockwise inference. However, these existing methods must be integrated with another online VAD system to help detect the presence of new speakers. The practical use of them is relatively difficult.

\begin{figure*}[t]
\centering
	\subfloat[Seq2Seq-TSVAD (offline-only)~\cite{cheng2023target}]{\includegraphics[height=1.65in]{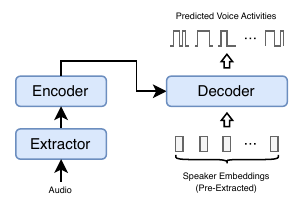}
	\label{fig_intro1}}
	\hfil
	\subfloat[S2SND (online/offline)]{\includegraphics[height=1.65in]{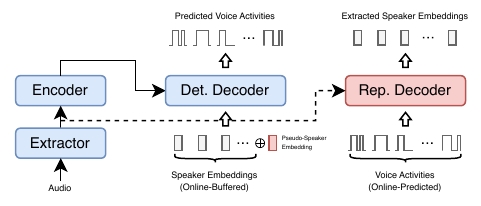}
	\label{fig_intro2}}
\caption{Overview of our speaker diarization frameworks from offline-only to online/offline scenarios: (a)~Previous Sequence-to-Sequence Target-Speaker Voice Activity Detection (Seq2Seq-TSVAD) framework; (b)~Newly proposed Sequence-to-Sequence Neural Diarization (S2SND) framework. \textit{Det.} and \textit{Rep.} denote the abbreviations of detection and representation, respectively. The red parts indicate the new modules added in the S2SND model compared to the Seq2Seq-TSVAD model.}
\label{fig_intro}
\end{figure*}

Fig.~\ref{fig_intro} illustrates the progress of our speaker diarization research. In our previous work~\cite{cheng2023target}, the Sequence-to-Sequence Target-Speaker Voice Activity Detection (Seq2Seq-TSVAD) framework has been proposed for offline-only speaker diarization. It mainly consists of three modules, shown in Fig.~\ref{fig_intro1}. First, the extractor obtains frame-level speaker embedding features from the raw audio. Next, the encoder processes long-term dependencies between frame-level features for the speaker diarization task. Finally, the decoder takes multiple speaker embeddings as reference information to predict the target-speaker voice activities, which has a one-to-one correspondence between the input order of speaker embeddings and the output order of voice activities. As a classical TSVAD-based method, the input speaker embeddings for the Seq2Seq-TSVAD model should be extracted by a prior diarization system, which restricts it from being an online system.

To tackle the above problem, this paper proposes a novel Sequence-to-Sequence Neural Diarization (S2SND) framework compatible with online and offline inference, shown in Fig.~\ref{fig_intro2}. The S2SND framework is still built upon the sequence-to-sequence architecture, with some improvements. First, The detection decoder works similarly to the single decoder in Fig.~\ref{fig_intro1}. Differently, we introduce a pseudo-speaker embedding to represent the unknown speaker without pre-extracted embedding, which is a kind of masked speaker prediction technique described in Sec.~\ref{sec:masked_speaker_prediction}. Second, we add a new representation decoder to take multiple target-speaker voice activities as reference information to predict speaker embeddings, which is a kind of target-voice speaker embedding extraction technique described in Sec.~\ref{sec:tvsee}. In this way, the S2SND framework can adopt partial speaker embeddings to predict complete voice activities and then extract the missed speaker embedding simultaneously. By traversing the input audio signal, the S2SND model can predict target-speaker voice activities of each coming audio block in real-time operation and progressively gather new speaker embeddings for the subsequent blocks. After the first-pass diarization, the collected speaker embeddings can also be used to re-decode the entire audio as an offline system.

Our proposed S2SND framework does not need unsupervised clustering or permutation-invariant training, making it fundamentally different from previous diarization systems. Therefore, we name it a new neural diarization approach. The contributions are summarized below.

\begin{enumerate}
\setlength{\itemsep}{0pt}
	\item We propose a novel masked speaker prediction method. One of the input speaker embeddings may be randomly erased during training. Then, the model learns to associate the output of the masked speaker with a learnable pseudo-speaker embedding, solving the one-to-one mapping problem between input speaker embeddings and output voice activities.
	\item We propose a novel target-voice speaker embedding extraction method. In contrast to the previous TSVAD method, it utilizes the predicted voice activities as reference information to extract target embeddings from the input audio further. The embedding space of speaker detection and representation is jointly learned.
	\item A simple but effective knowledge distillation strategy is developed to explore the potential of our proposed method when meeting large-scale data. We evaluate our approach on several widely-used datasets, outperforming previous state-of-the-art results in various online and offline evaluation settings.
	\item The designed framework combines characteristics of both EEND and TSVAD methods. It is not only clustering-free and PIT-free, but also can utilize the end-to-end neural network to discover possible unknown speakers. Meanwhile, the use of target embeddings maintains recognized speaker identities consistent across different blocks in long audio, which is usually the advantage of TSVAD-based methods.
	\end{enumerate}

\section{Related Works}

\subsection{Offline Diarization}

The cascaded speaker diarization consists of several components. 1) Voice Activity Detection (VAD)~\cite{chang2018temporal} removes non-speech regions from the audio. 2) Speech regions are divided into shorter segments~\cite{hruz2017convolutional,sell2018diarization}. 3) Speaker embeddings (e.g., i-vectors~\cite{dehak2010front}, x-vectors~\cite{snyder2018x}) are extracted from the speech segments and clustered into different identities by K-Means~\cite{wang2018speaker}, AHC~\cite{sell2018diarization}, SC~\cite{lin2019lstm}, or others. 4) Post-processing techniques for overlapped speech regions can be optionally implemented~\cite{landini2020but,bredin2021end}. The number of output speakers is determined by the clustering algorithms.

End-to-End Neural Diarization (EEND)~\cite{fujita2019end_1,fujita2019end_2} predicts multiple speakers' voice activities by formulating the diarization problem as a multi-label classification task. The original EEND models have a fixed number of output speakers restricted by their network architecture. Although using Encoder-Decoder based Attractor (EDA)~\cite{horiguchi2020end,horiguchi2022encoder} can infer the variable number of speakers. In practice, the number of output speakers is still capped by the training data~\cite{takashima2021end}. To solve this problem, integrating the end-to-end and clustering approach is a promising direction. For example, EEND-vector clustering (EEND-VC)~\cite{kinoshita2021integrating,kinoshita2021advances,kinoshita2022tight} deploys an EEND model for shortly divided audio blocks and addresses the inter-block speaker permutation ambiguity by clustering of speaker embeddings. EEND-GLA~\cite{horiguchi2021towards,horiguchi2022online} computes local attractors for each short block and determines speaker correspondence based on similarities between inter-block attractors. Also, several extensions of EEND are proposed from the aspects of network architecture~\cite{rybicka2022end,fujita2023neural,landini2024diaper}, objective function design~\cite{jeoung2023improving,palzer2024improving}, self/semi-supervised learning~\cite{dissen2022self,takashima2021semi}, and so on.

Target-Speaker Voice Activity Detection (TSVAD)~\cite{medennikov2020target} is also effective. It relies on a prior diarization system to extract each speaker's acoustic footprint (i-vector) as enrollment. Then, the TSVAD model uses speech features (e.g., MFCC) and extracted i-vectors to output target speaker voice activities according to the enrollment order. Later, He et al.~\cite{he2021target} adapt the model to handle a variable number of speakers by setting a maximum speaker limit and producing null voice activities for zero-padded ones. Sequential models (e.g., LSTM~\cite{Cheng_2023} and Transformer~\cite{wang2023target}) are implemented on the speaker dimension of model input to manage a variable number of speakers. To explore more discriminative speaker embeddings as an alternative to i-vector, Wang et al.~\cite{wang2022similarity} replace the front-end of the TSVAD model with a pre-trained extractor tailored for frame-level x-vectors. This modification demonstrates superior performance than a simple swap of i-vectors for x-vectors in early attempt~\cite{medennikov2020target}. Furthermore, the TSVAD framework has been investigated in various aspects (e.g., multi-channel signal~\cite{wang2022cross}, multi-modal system~\cite{cheng2023whu,cheng2024multi,jiang2024target}, joint inference with ASR~\cite{wang2024joint}, generative approach~\cite{chen2024flow}).

In addition, several studies have explored using voice activity information to guide speaker embedding extraction for the downstream tasks, including both clustering-based and TSVAD-based diarization systems~\cite{10887711,he2023ansd}. However, these methods treat embedding extraction as a separate stage without jointly optimizing it with diarization objectives. Also, they are inherently designed for offline inference and thus cannot be directly extended to low-latency online scenarios.

\subsection{Online Diarization}

In an online scenario, the diarization system must make continual decisions on each audio frame while the conversation continues. This paradigm is crucial for low-latency applications such as real-time conversation transcription.

To extend cascaded methods to online inference, all built-in modules (e.g., voice activity detection, speech segmentation, speaker embedding extraction) must be executed in real time. Several techniques (e.g., UIS-RNN~\cite{zhang2019fully}, UIS-RNN-SML~\cite{fini2020supervised}) replace the speech segmentation and speaker clustering with supervised neural networks. As the most critical component,  online speaker clustering attracts much research interest, e.g., modified clustering~\cite{dimitriadis2017developing,coria2021overlap}, PLDA-scoring~\cite{sholokhov2023probabilistic}, and clustering guided embedding extractor training~\cite{chen2024interrelate}. However, their time complexity will increase with the number of speech segments, resulting in inadequate performance for long audio.

The extension of end-to-end approaches to online diarization are broadly divided into two directions. The first is to train models that can convey information during block-wise or frame-wise inference to address the speaker permutation ambiguity. For instance, BW-EDA-EEND~\cite{han2021bw} adopts Transformer-XL~\cite{dai2019transformer} with recursive hidden states to take block-wise inputs, where the hidden states obtained from the previous blocks are used to generate attractors of the current block. Liang et al.~\cite{liang2024frame} propose the frame-wise online EEND (FS-EEND) to adaptively update speaker attractors frame by frame, which has a lower inference latency. In this direction, online diarization models are easily optimized in a fully end-to-end manner. However, independent network architectures are required rather than offline diarization models. If both offline and online diarization models are needed, the deployment costs will be largely increased. The second direction is to modify offline models for online inference. Speaker-tracing buffer (STB)~\cite{xue2021online_1, xue2021online_2} is proposed to maintain the preceding results of EEND models during online inference. It makes the order of output speakers consistent without changing the network architecture. On top of this direction, EEND-GLA~\cite{horiguchi2021towards,horiguchi2022online} further integrates local and global attractors with STB for online inference, achieving state-of-the-art performance on multiple datasets. It is reported that STB can minimize the inference latency using a small block size and outperform BW-EDA-EEND~\cite{xue2021online_2,horiguchi2022online}. Nonetheless, this approach demands extra computations because every past frame in the buffer must be re-computed for each new block.

\begin{figure*}[t]
\centering
  \includegraphics[width=\linewidth]{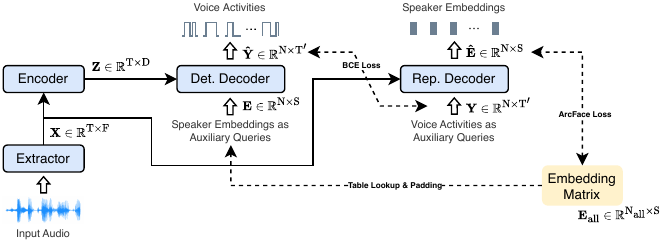}
  \caption{The Sequence-to-Sequence Neural Diarization (S2SND) framework. \textit{Det.} and \textit{Rep.} denote the abbreviations of detection and representation, respectively.}
  \label{fig:framework}
\end{figure*}

On the success of offline TSVAD methods~\cite{wang2021ustc,wang2021dku,wang2022dku,cheng2023dku}, diverting the TSVAD framework for online inference is also a promising direction. In offline scenarios, TSVAD methods usually serve as post-processing to refine cascaded diarization results~\cite{medennikov2020target}. After obtaining target-speaker embedding from the initial stage, TSVAD models process audio signals block-wise. This property implies that TSVAD models are naturally adapted to online inference if target-speaker embeddings are acquired in real time. Therefore, Wang et al.~\cite{wang22j_interspeech} firstly present the online TSVAD framework and then adapt it to multi-channel data~\cite{wang2023end}. Chen et al.~\cite{chen2024enhancing} design a dictionary learning module across different frequency bands in multi-channel data to reduce the inference cost. Nevertheless, two problems prevent existing online TSVAD methods~\cite{wang22j_interspeech,wang2023end,chen2024enhancing} from practical use. 1) They assume the input audio to contain at least one active speaker at all times. During inference, unenrolled speakers are determined once the models do not detect any active voice activity using enrolled speaker embeddings. Thus, an additional VAD module is required to remove all the silent regions in the input audio. The VAD errors might severely impact the final system output. This kind of bypass approach does not fundamentally solve the problem of new speaker detection. 2) They utilize local speaker labels within each recording to optimize target embedding extraction, where the power of global speaker modeling is not fully exploited. In contrast, current advanced speaker verification techniques are mainly based on unique speaker identities over the whole training set~\cite{wang2024overview}.

Notably, the concepts of TSVAD and EEND families are becoming closer. In early TSVAD systems~\cite{medennikov2020target,wang2022similarity,cheng2023target}, speaker embeddings are typically acoustic footprints extracted by speaker verification models (e.g., i-vectors~\cite{dehak2010front}, x-vectors~\cite{snyder2018x}). Then, works of~\cite{wang22j_interspeech, wang2023end} turn to generate speaker embeddings within TSVAD models. On the other hand, the attractors used in EEND systems~\cite{horiguchi2020end,horiguchi2022encoder} are also a kind of local speaker embeddings within each audio block. Recent studies of~\cite{liang2024frame,horiguchi2022online} begin to constrain the speaker similarity of attractors across different audio blocks. Obviously, using a set of embedding vectors to represent speaker identities has been widely adopted with different terminologies. Therefore, in this work, we aim to take advantage of both EEND and TSVAD methods to propose the new S2SND framework, achieving state-of-the-art performance on various multi-scenario datasets.

\section{Sequence-to-Sequence Neural Diarization}

\subsection{Architecture} 

The proposed S2SND framework takes the sequence-to-sequence architecture used in our previous offline method~\cite{cheng2023target} with several modifications, shown in Fig.~\ref{fig:framework}. 

\subsubsection{Extractor}
The ResNet-based~\cite{he2016deep} model is adopted as the front-end extractor. The audio signal is firstly transformed into log Mel-filterbank energies. Then, it is fed into the extractor with segmental statistical pooling (SSP)~\cite{wang2022similarity} to obtain frame-level speaker embeddings $ \mathbf{X} \in \mathbb{R}^\mathrm{T\times F}$, where $T$ and $F$ denote the length and dimension of extracted feature sequence. An additional linear layer is employed to align the output dimension $F$  with the input dimension of subsequent encoder and decoder modules, omitted to plot for clarity. This process converts raw audio signals into a sequence of neural network-based features.

\subsubsection{Encoder}
The Conformer-based~\cite{gulati2020conformer} model is employed as the encoder to process the frame-level speaker embeddings. The input feature sequence is firstly added with sinusoidal positional encodings~\cite{vaswani2017attention} and then fed into the encoder to obtain output feature sequence $ \mathbf{Z} \in \mathbb{R}^\mathrm{T\times D}$, where $D$ is the attention dimension used in the encoder. This process further takes long-term dependencies between frame-level speaker embeddings for the diarization task.

\subsubsection{Decoder}

\begin{figure}[t]
\centering
  \includegraphics[width=\linewidth]{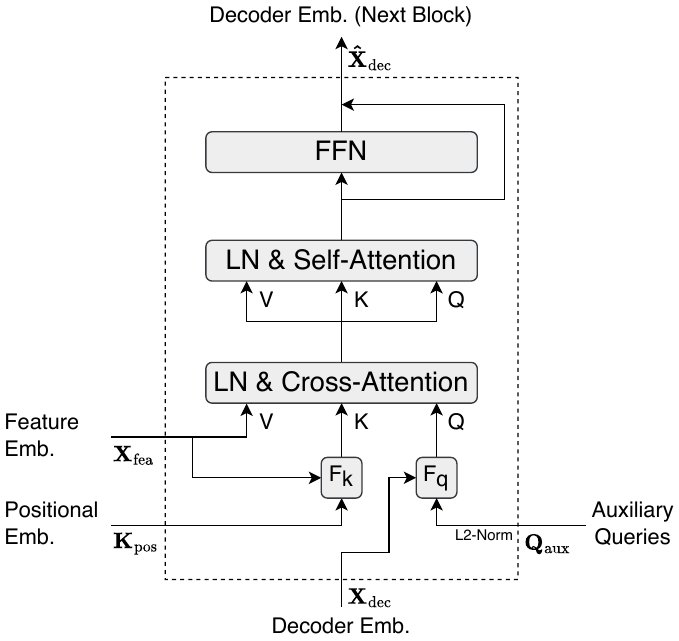}
  \caption{The structure of the modified Speaker-wise Decoder. For clarity, the residual connections between attention layers are omitted from the plot. The abbreviation of \textit{LN} refers to the layer normalization applied before the QKV inputs for attention modules.}
  \label{fig:decoder}
\end{figure}

The decoder block retains the main layout of the original Speaker-wise Decoder (SW-D)~\cite{cheng2023target} with a few changes, shown in Fig.~\ref{fig:decoder}. There are four parts of the input. First, the feature embeddings come from the extractor output $\mathbf{X}$ or encoder output $\mathbf{Z}$. Second, the positional embeddings are the same as those used in the encoder. Third, as a decoder module is usually composed of several basic decoder blocks stacked together, the input decoder embeddings at the first block are initialized by zeros and then processed by the following blocks. After the last output block, a simple linear transformation is adopted to obtain the desirable output dimension. Lastly, the auxiliary queries play the role of reference information in multi-speaker tasks, which can be either target-speaker embeddings or voice activities. The detailed design of the decoder block is described below.

\begin{itemize}
	\item The cross-attention layer is placed before the self-attention layer. As the input embeddings of the first decoder block are initialized as zeros, there is no useful information for the self-attention layer at first.
	\item We define $\mathrm{F}_\mathrm{q}(\cdot)$ and $\mathrm{F}_\mathrm{k}(\cdot)$ to denote the fusion operations for input queries and keys, respectively. Let $N$ denote the preset maximum number of speakers that the model can handle simultaneously. $\mathbf{X}_\mathrm{dec} \in \mathbb{R}^\mathrm{N \times D}$ and $\mathbf{Q}_\mathrm{aux} \in \mathbb{R}^\mathrm{N \times D^{\prime}}$ represent the decoder embeddings and auxiliary queries, respectively. The $\mathrm{F}_\mathrm{q}(\cdot)$ operation is described as:
	\begin{align}
		\mathbf{Q} =  \mathbf{X}_\mathrm{dec} + \mathrm{Linear}_{\mathbb{R}^\mathrm{{D^{\prime}}} \rightarrow \mathbb{R}^\mathrm{D}}(\mathbf{Q}_\mathrm{aux}) / \sqrt{D},
	\end{align} 
	where the linear transformation is deployed to align the dimension of queries and keys with the weight factor $1/\sqrt{D}$. Similarly, let $\mathbf{X}_\mathrm{fea} \in \mathbb{R}^\mathrm{T \times D}$ and $\mathbf{K}_\mathrm{pos} \in \mathbb{R}^\mathrm{T \times D^{\prime}}$ represent the feature embeddings and positional embeddings with the length of $T$. The $\mathrm{F}_\mathrm{k}(\cdot)$ operation is described as:
	\begin{align}
		\mathbf{K} = \mathbf{X}_\mathrm{fea} + \mathrm{Linear}_{\mathbb{R}^\mathrm{D^{\prime}} \rightarrow \mathbb{R}^\mathrm{D}}(\mathbf{K}_\mathrm{pos}) / \sqrt{D}.  
	\end{align}
	The fused queries $\mathbf{Q}$ and keys $\mathbf{K}$ are fed into the cross-attention layer with a Pre-LayerNorm method. Compared with the previous concatenation fusion, this additive fusion is more straightforward without expanding the output dimension of queries and keys.
	\item If the input auxiliary queries are speaker embeddings, they must be $L2$-normalized to lie in a hypersphere. Otherwise, if the input auxiliary queries are voice activities, they do not need to undergo any normalization.
\end{itemize}

Based on the same structure of the decoder block, we introduce two decoders responsible for different functions. Let $\mathbf{E} \in \mathbb{R}^\mathrm{N\times S}$ denote the given speaker embeddings with the number of $N$ and the dimension of $S$. The ground truth of their target voice activities is denoted as a binary matrix $ \mathbf{Y}  \in \left \{0, 1 \right \}^\mathrm{N\times T^{\prime}}$, where $y_{n, t^{\prime}}$ represents the speaking existence of the $n$-th speaker at time $t^{\prime}$. The detection decoder utilizes encoder output $\mathbf{Z}$ as feature embeddings and speaker embeddings $\mathbf{E}$ as auxiliary queries to obtain the predicted voice activities $\mathbf{\hat{Y}} \in \left \{0, 1 \right \}^\mathrm{N\times T^{\prime}}$. In contrast, the representation decoder utilizes extractor output $\mathbf{X}$ as feature embeddings and voice activities $ \mathbf{Y}$ as auxiliary queries to obtain the extracted speaker embeddings $\mathbf{\hat{E}} \in \mathbb{R}^\mathrm{N\times S}$. Two decoders perform inverse tasks to predict target-speaker voice activities and extract speaker embeddings simultaneously.

\subsection{Training Process}
\label{sec:training}

The ground truth of $\mathbf{Y}$ is obtained from the adopted dataset during training. However, $\mathbf{E}$ is not directly available because the embedding space must be learned by neural networks. To overcome this problem, we adopt a learnable embedding matrix $\mathbf{E}_\mathrm{all} \in \mathbb{R}^\mathrm{N_{all} \times S}$ as the target embeddings of all speakers in the training data. $N_\mathrm{all}$ and $S$ represent the total number of speakers and embedding dimension, respectively. Each row vector denotes one specific speaker embedding, which is randomly initialized as a unit vector with a magnitude (norm) of 1. Given $n \le N_\mathrm{all}$, the $n$-th target-speaker embedding in the training data is obtained by $\mathbf{E}_\mathrm{all}(n,:)$. Meanwhile, the $n$-th speaker label in the training data is denoted by a $N_\mathrm{all}$-dim one-hot vector with zeros everywhere except its $n$-th value will be 1. During training, given an input audio block with $N_\mathrm{loc}$ speaker labels $\mathbf{S}_\mathrm{loc} \in \left ( 0,1 \right )^\mathrm{N_{loc} \times N_{all}}$, the input speaker embeddings for the detection decoder are obtained by $\mathbf{S}_\mathrm{loc} \cdot \mathbf{E}_\mathrm{all} \in \mathbb{R}^\mathrm{N_{loc} \times S}$,  which is a simple table look-up operation using matrix multiplication. Also, $\mathbf{E}_\mathrm{all}$ is used as the training objectives of output speaker embeddings from the representation decoder. We propose the following approaches to jointly optimize $\mathbf{E}_\mathrm{all}$ with the whole diarization model.

\subsubsection{Masked speaker prediction}
\label{sec:masked_speaker_prediction}

The masked language modeling (MLM) technique has been validated in natural language processing~\cite{devlin2018bert}, conducted by randomly masking some words in the input text and then training the model to predict the masked words. Similarly, we introduce a masked speaker prediction method into speaker diarization. During training, one of the input speaker embeddings for each audio block will be randomly masked. The model learns to identify whether there is a person speaking in the audio block but without the given speaker embedding. To achieve this goal, two padding strategies are implemented.

The first strategy is to pad the input speaker embeddings $\mathbf{E}$ using a learnable pseudo-speaker embedding $\mathbf{e}_\mathrm{pse} \in \mathbb{R}^\mathrm{S}$, where $\mathbf{e}_\mathrm{pse}$ is initialized with zeros and optimized during training. For each training data, a probability is 0.5 that one existing speaker label $\mathbf{s}_{n} \in \mathbf{S}$ will be randomly selected as the masked one. Accordingly, the speaker embedding $\mathbf{e}_{n}$ will be removed from $\mathbf{E}$ and the ground-truth of target voice activities $\mathbf{y}_{n} \in \mathbf{Y}$ will be re-assigned to the output of pseudo-speaker embedding. In this way, the model is trained to utilize the pseudo-speaker embedding to capture any unenrolled speaker's voice activities.

The second strategy is to pad the input speaker embeddings $\mathbf{E}$ using a learnable non-speech embedding $\mathbf{e}_\mathrm{non} \in \mathbb{R}^\mathrm{S}$, where $\mathbf{e}_\mathrm{non}$ is initialized with zeros and optimized during training. We define speaker capacity as the maximum number of speakers (embeddings) that the model can process simultaneously. By setting the speaker capacity to a relatively large value $N$, the pseudo-speaker embedding $\mathbf{e}_\mathrm{pse}$ accounts for one, and there are also $N_\mathrm{loc}$ existing speaker embeddings. As usually $(N_\mathrm{loc}+1) \ll N$, the left $N - N_\mathrm{loc} - 1$ vacancies will be randomly filled up with 50\% non-speech embeddings and 50\% speaker embeddings from who are not appearing in the current audio block. Accordingly, their ground-truth voice activities are silent. In this way, the input dimension of a mini-batched training data is aligned, and the model is trained to distinguish valid and invalid speaker embeddings for the given audio block and how to assign target voice activities to the corresponding speakers.

Finally, the output $\mathbf{\hat{Y}}$ from the detection decoder is optimized to minimize its binary cross-entropy (BCE) loss with $\mathbf{Y}$, which is described as follows:
\begin{align}
\begin{split}
\mathcal{L}_{\mathrm{bce}} = - \frac{1}{N\times T^{\prime}} \sum_{n=1}^{N} \sum_{t^{\prime}=1}^{T^{\prime}} 
& \left[y_{n,t^{\prime}} \log(\hat{y}_{n,t^{\prime}}) + \right. \\ 
& \left. (1 - y_{n,t^{\prime}}) \log(1 - \hat{y}_{n,t^{\prime}}) \right],
\end{split}
\label{eq:bce}
\end{align}
\noindent where $\hat{y}_{n,t^{\prime}} = \mathbf{\hat{Y}}(n,t^{\prime})$ is the predicted speaking probability of the $n$-th speaker at time $t^{\prime}$. And $y_{n,t^{\prime}} = \mathbf{Y}(n,t^{\prime})$ is its ground-truth label.

\begin{figure*}[t]
\centering
  \includegraphics[width=\linewidth]{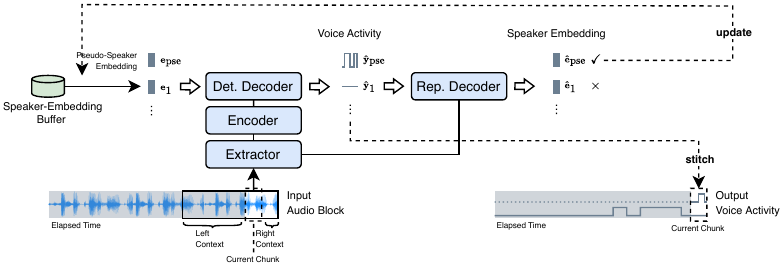}
  \caption{The inference diagram of the Sequence-to-Sequence Neural Diarization (S2SND) framework. \textit{Det.} and \textit{Rep.} denote the abbreviations of detection and representation, respectively.}
  \label{fig:inference}
\end{figure*}

\subsubsection{Target-voice speaker embedding extraction}
\label{sec:tvsee}
 
As speaker embeddings can be used as reference information to extract target speaker voice activities, why can't voice activities be used as reference information to extract target speaker embeddings from multi-talker audio signals? Although a voice activity pattern may be the same between two speakers in a short window, it is not a big issue because the single-speaker speech usually occupies most of the time in real conversational data. Following this idea, we propose the target-voice speaker embedding extraction method, an inverse function of target-speaker voice activity detection.

The ArcFace~\cite{deng2019arcface} loss is employed between $\mathbf{\hat{E}}$ and the embedding matrix $\mathbf{E}_\mathrm{all}$, which is described as follows:
\begin{align}
\mathcal{L}_{\mathrm{arc}} = \frac{1}{N} \sum_{n=1}^{N} -\log\frac{e^{\alpha \cdot \cos(\theta_{n}+m)}}{e^{\alpha \cdot \cos(\theta_{n}+m)}+\sum_{i=1,i\neq \phi(n)}^{N_{all}}e^{\alpha \cdot \cos\theta_{i}}}.
\label{eq:arcface} 
\end{align}

\noindent In this formula, $n$ is the local speaker index for the current model output within the maximum speaker capacity, where $1 \le n \le N$. Let $\phi(n)$ denote the mapping from the local speaker index $n$ to its corresponding global speaker index within all training data, where $1 \le \phi(n) \le N_\mathrm{all}$. In code implementation, $\phi$ could be easily recorded when preparing training data. Then, $\theta_{n}$ is the angle between the $n$-th extracted speaker embedding $\mathbf{\hat{e}}_{n} \in \mathbf{\hat{E}}$ and its target embedding $\mathbf{e}_{\phi(n)} \in \mathbf{E}_\mathrm{all}$. $\theta_{i}$ is the angle between the $n$-th extracted speaker embedding $\mathbf{\hat{e}}_{n} \in \mathbf{\hat{E}}$ and its non-target embedding $\mathbf{e}_{i} \in \mathbf{E}_\mathrm{all}$, where $1 \le i \le N_\mathrm{all}$ and $i\neq \phi(n)$ controls that $\theta_{i}$ is only implemented on negative pairs in contrastive learning. $\alpha$ and $m$ are the re-scale factor and additive angular margin penalty, respectively.

The total training loss is the sum of $\mathcal{L}_{\mathrm{bce}}$ in Eq.~\ref{eq:bce} and $\mathcal{L}_{\mathrm{arc}}$ in Eq.~\ref{eq:arcface}. Using a learnable embedding matrix as the bridge between the built-in decoders, the embedding space of speaker detection and representation is jointly optimized in an end-to-end manner.

\subsection{Inferring Process}
\label{sec:inferring}

Fig.~\ref{fig:inference} demonstrates the inference diagram of our proposed S2SND framework. Once the training is finished, the embedding matrix will no longer be needed. Instead, a speaker-embedding buffer is initialized as an empty dictionary to store speaker embeddings extracted during inference. Then, the model processes the input audio block by block and progressively updates the speaker-embedding buffer.

\subsubsection{Data preparation}

The input audio is cut into blocks of fixed length $L$, where $L$ should be identical to the preset during training. To reduce the latency of model output, we introduce a smaller unit: chunk. As shown in the left part of Fig.~\ref{fig:inference}, each input audio block contains three regions: left context, current chunk, and right context. The chunk length is set to $L_\mathrm{chunk}$, representing the period corresponding to each inference step. The left and right context lengths are set to $L_\mathrm{left}$ and $L_\mathrm{right}$, respectively. A sliding window method is applied to move the current chunk on the audio stream with the chunk shift equal to the chunk length. For each inference, the model takes the input audio block containing the current chunk and its contexts as long as $L = L_\mathrm{left}+L_\mathrm{chunk}+L_\mathrm{right}$. The absence of left context is padded with zeros at the beginning of the inference until the acquired audio signal is available to compose an entire block. Furthermore, since acquiring the right context needs to await an extra period, the algorithmic latency of model inference should be the sum of $L_\mathrm{chunk}$ and $L_\mathrm{right}$.

Assume that speaker capacity is set to $N$ during training and $N_\mathrm{loc}$ identities are currently enrolled in the speaker-embedding buffer. The input speaker embeddings for the detection decoder consist of three different sources. The first part is always kept for the pseudo-speaker embedding $\mathbf{e}_\mathrm{pse} \in \mathbb{R}^\mathrm{S}$. The second part consists of the target embeddings enrolled in the current speaker-embedding buffer, denoted as $\mathbf{E}_\mathrm{buf} = \left [ \mathbf{e}_{1}\ \mathbf{e}_{2}\ \cdots\ \mathbf{e}_\mathrm{N_{loc}} \right ]^{\top} \in \mathbb{R}^\mathrm{N_{loc} \times S}$. The third part is padded by the non-speech embedding $\mathbf{e}_\mathrm{non} \in \mathbb{R}^\mathrm{S}$ with the number of $N - N_\mathrm{loc} - 1$, denoted as $\mathbf{E}_\mathrm{non} = \left [ \mathbf{e}_\mathrm{non}\ \mathbf{e}_\mathrm{non}\ \cdots\ \mathbf{e}_\mathrm{non} \right ]^{\top} \in \mathbb{R}^\mathrm{(N-N_{loc}-1) \times S}$. Overall, the total input speaker embeddings are concatenated by $\mathbf{E} = \left [ \mathbf{e}_\mathrm{pse} \ \mathbf{E}_\mathrm{buf}^{\top} \ \mathbf{E}_\mathrm{non}^{\top} \right ]^{\top} \in \mathbb{R}^\mathrm{N \times S}$. Therefore, the dimension of input speaker embeddings during inference is consistent with the preset during training.

\subsubsection{Decoding Procedure}

The first decoding stage takes the given speaker embeddings as reference information to predict multiple speakers' voice activities from the detection decoder. As the input order of speaker embeddings determines the output order of target voice activities, the predicted target-speaker voice activities also have three parts. Let $T^{\prime}$ indicate the number of timestamps in a given speaker's prediction. The first part is $\mathbf{\hat{y}}_\mathrm{pse} \in \mathbb{R}^\mathrm{T^{\prime}} $, which represnets the predicted result corresponding to the pseudo-speaker embedding $\mathbf{e}_\mathrm{pse}$. The second part is made of the predicted results corresponding to the buffered embeddings $\mathbf{E}_\mathrm{buf}$, denoted as $\mathbf{\hat{Y}}_\mathrm{buf} = \left [ \mathbf{\hat{y}}_{1}\ \mathbf{\hat{y}}_{2}\ \cdots\ \mathbf{\hat{y}}_\mathrm{N_{loc}} \right ]^{\top} \in \mathbb{R}^\mathrm{N_{loc} \times T^{\prime}}$. The third part is padded by the predicted results corresponding to the non-speech embeddings $\mathbf{E}_\mathrm{non}$, denoted as $\mathbf{\hat{Y}}_\mathrm{non} = \left [ \mathbf{\hat{y}}_\mathrm{non}\ \mathbf{\hat{y}}_\mathrm{non}\ \cdots\ \mathbf{\hat{y}}_\mathrm{non} \right ]^{\top} \in \mathbb{R}^\mathrm{(N-N_{loc}-1) \times T^{\prime}}$. Overall, the total predicted target-speaker voice activities are concatenated by $\mathbf{\hat{Y}} = \left [ \mathbf{\hat{y}}_\mathrm{pse} \ \mathbf{\hat{Y}}_\mathrm{buf}^{\top} \ \mathbf{\hat{Y}}_\mathrm{non}^{\top} \right ]^{\top} \in \mathbb{R}^\mathrm{N \times T^{\prime}}$. Furthermore, $\mathbf{\hat{Y}}_\mathrm{non}$ are invalid results because they belong to padded contents to maintain the fixed dimension of predicted target-speaker voice activities.

The second decoding stage takes the predicted voice activities as reference information to extract multiple speakers' embeddings from the representation decoder. Similarly, the input order of voice activities determines the output order of target speaker embeddings. First, $\mathbf{\hat{e}}_\mathrm{pse} \in \mathbb{R}^\mathrm{S}$ represents the extracted result corresponding to $\mathbf{\hat{y}}_\mathrm{pse}$. Second, $\mathbf{\hat{E}}_\mathrm{buf} = \left [ \mathbf{\hat{e}}_{1}\ \mathbf{\hat{e}}_{2}\ \cdots\ \mathbf{\hat{e}}_\mathrm{N_{loc}} \right ]^{\top} \in \mathbb{R}^\mathrm{N_{loc} \times S}$ denotes the extracted results corresponding to $\mathbf{\hat{Y}}_\mathrm{buf}$. Third, $\mathbf{\hat{E}}_\mathrm{non} = \left [ \mathbf{\hat{e}}_\mathrm{non}\ \mathbf{\hat{e}}_\mathrm{non}\ \cdots\ \mathbf{\hat{e}}_\mathrm{non} \right ]^{\top} \in \mathbb{R}^\mathrm{(N-N_{loc}-1) \times S}$ denotes the extracted results corresponding to $\mathbf{\hat{Y}}_\mathrm{non}$. Overall, the total extracted target-speaker embeddings are concatenated by three parts, denoted as $\mathbf{\hat{E}} = \left [ \mathbf{\hat{e}}_\mathrm{pse} \ \mathbf{\hat{E}}_\mathrm{buf}^{\top} \ \mathbf{\hat{E}}_\mathrm{non}^{\top} \right ]^{\top} \in \mathbb{R}^\mathrm{N \times S}$.

Furthermore, let $\mathbf{\hat{y}} = [y_{1}, y_{2}, \dots, y_{T^{\prime}}]$ indicates the predicted voice activities of a given speaker in $\mathbf{\hat{Y}}$, we define the operation $W: \mathbb{R}^\mathrm{T^{\prime}} \rightarrow \mathbb{R}, \mathrm{W}(\mathbf{\hat{y}}) = \sum_{t^{\prime}=1, t^{\prime} \notin \mathbf{Overlap}}^{T^{\prime}} \hat{y}_{t^{\prime}}$ to count the non-overlapped speaking time in the given $\mathbf{\hat{y}}$. As the quality of embedding extraction may be easily affected by each speaker's active speaking time and overlapping status, the longer single-speaking time for each speaker usually results in better embedding extraction, which can be used as an additional embedding weight. Applying the function $W$ to each predicted target-speaker voice activity in $\mathbf{\hat{Y}}$, the weight of each extracted target-speaker embedding is calculated one by one. First, $\hat{w}_\mathrm{pse} \in \mathbb{R}$ represents the weight corresponding to $\mathbf{\hat{e}}_\mathrm{pse}$. Second, $\mathbf{\hat{w}}_\mathrm{buf} = [\hat{w}_{1}, \hat{w}_{2}, \dots, \hat{w}_\mathrm{N_{loc}}]^{\top} \in \mathbb{R}^\mathrm{N_{loc}}$ denotes the weights corresponding to $\mathbf{\hat{E}}_\mathrm{buf}$. Third, $\mathbf{\hat{w}}_\mathrm{non} = [\hat{w}_\mathrm{non}, \hat{w}_\mathrm{non}, \dots, \hat{w}_\mathrm{non}]^{\top} \in \mathbb{R}^\mathrm{N-N_{loc}-1}$ denotes the weights corresponding to $\mathbf{\hat{E}}_\mathrm{non}$. Overall, the total weights of extracted target-speaker embeddings are concatenated by three parts, denoted as $\mathbf{\hat{w}} = \left [ \hat{w}_\mathrm{pse} \ \mathbf{\hat{w}}_\mathrm{buf}^{\top} \ \mathbf{\hat{w}}_\mathrm{non}^{\top} \right ]^{\top} \in \mathbb{R}^\mathrm{N}$.

\begin{figure}[t]
\centering
  \includegraphics[width=\linewidth]{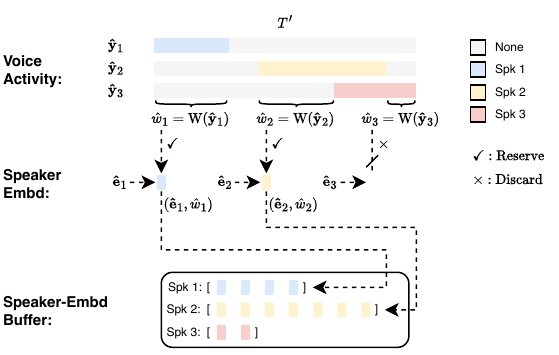}
  \caption{Updating strategy of the speaker-embedding buffer.}
  \label{fig:buffer}
\end{figure}

%########################################################################
\begin{algorithm*}[!t]
\caption{Pseudocode of online inference in the Python-like style.}
\label{alg:code} 
\definecolor{codeblue}{rgb}{0.25,0.5,0.5}
\lstset{
  backgroundcolor=\color{white},
  basicstyle=\fontsize{8pt}{8pt}\ttfamily\selectfont,
  columns=fullflexible,
  breaklines=true,
  captionpos=b,
  commentstyle=\fontsize{8pt}{8pt}\color{codeblue},
  keywordstyle=\fontsize{8pt}{8pt},
}

\begin{lstlisting}[language=python]
"""
- Extractor(), Encoder(), Det_Decoder(), Rep_Decoder(): neural network modules in S2SND models
- W(): calculating embedding weight  
Inputs
- blocks: a sequence of input audio blocks 
- e_pse/e_non: pseudo-speaker/non-speech embedding
- tau_1/tau_2: threshold for pseudo-speaker/enrolled-speaker embedding weight
- lc/lr: number of output VAD frames belonging to the current chunk / right context
- N: speaker capacity
- S: embedding dimension
Outputs
- dia_result: predicted target-speaker voice activities
- emb_buffer: extracted speaker embeddings
"""

dia_result = {} # initial diarization result
emb_buffer = {} # initial speaker-embedding buffer
num_frames = 0  # number of predicted VAD frames

for audio_block in blocks: # load the next audio block
    emb_list = [e_pse] # initialize input speaker embedding list & put pseudo-speaker embedding
    spk_list = [len(emb_buffer)+1] # initialize input speaker labels & create new speaker label
    
    for spk_id in emb_buffer.keys(): # obtain each enrolled target-speaker embedding
        e_sum = torch.zeros(S) # embedding vector, shape: S 
        w_sum = 0 # embedding weight, scalar
        for e_i, w_i in emb_buffer[spk_id]: 
            e_sum += w_i*e_i
            w_sum += w_i
        emb_list.append(e_sum/w_sum) # append weighted speaker embedding
        spk_list.append(spk_id) # append speaker label
        
    while len(emb_list) < N: # pad input embeddings to tensor with the length of N
        emb_list.append(e_non)
    emb_tensor = torch.stack(emb_list)
    
    X = Extractor(audio_block) # forward extractor, output shape: T x F
    X_hat = Encoder(X) # forward encoder, output shape: T x D
    Y_hat = Det_Decoder(X_hat, emb_tensor) # forward detection decoder, output shape: N x T'
    E_hat = Rep_Decoder(X, Y_hat) # forward representation decoder, output shape: N x S
    
    y_pse = Y_hat[0] # predicted pseudo-speaker voice activity, shape: T'
    e_pse = E_hat[0] # extracted pseudo-speaker embedding, shape: S
    w_pse = W(y_pse) # embedding weight: scalar
    if w_pse > tau_1: 
        elapsed_y = torch.zeros(num_frames) # create the elapsed result as zeros
        current_y = y_pse[-(lc+lr):-lr] # cut the current chunk result from the block output
        new_id = spk_list[0] # get speaker id
        dia_result[new_id] = torch.cat(elapsed_y, current_y) # store diarization result
        emb_buffer[new_id] = [(e_pse, w_pse)] # store embedding-weight pair
        
    for n in range(1, len(S)):
        y_n = Y_hat[n] # predicted enrolled voice activity, shape: T'
        e_n = E_hat[n] # extracted enrolled embedding, shape: S
        w_n = W(y_n) # embedding weight, scalar
        spk_id = spk_list[n] # get speaker id
        dia_result[spk_id] = torch.cat(dia_result[spk_id], y_n[-(lc+lr):-lr]) # stitch diarization result
        if w_n > tau_2:  
            emb_buffer[spk_id].append((e_n, w_n)) # append embedding-weight pair
            
    num_frames += lc # update
\end{lstlisting}
\end{algorithm*}
%########################################################################

We adopt two thresholds denoted as $\tau_{1}$ and $\tau_{2}$, respectively. If $\hat{w}_\mathrm{pse} > \tau_{1}$, it means that an unenrolled speaker is detected and the extracted embedding is qualified to be reserved. The results of $\mathbf{\hat{y}}_\mathrm{pse}$ and $\mathbf{\hat{e}}_\mathrm{pse}$ will be assigned a new speaker label. Otherwise, $\mathbf{\hat{y}}_\mathrm{pse}$ and $\mathbf{\hat{e}}_\mathrm{pse}$ will be discarded as invalid results. Also, each $\mathbf{\hat{e}}_{n} \in \mathbf{\hat{E}}_\mathrm{buf}$ represents the extracted target-speaker embedding corresponding to the target-speaker voice activity $\mathbf{\hat{y}}_{n}$, where $1 \le n \le N_\mathrm{loc}$. If $\hat{w}_{n} > \tau_{2}$, the results of $\mathbf{\hat{y}}_{n}$ and $\mathbf{\hat{e}}_{n}$ will be reserved. Otherwise, $\mathbf{\hat{e}}_{n}$ will be discarded to prevent the unreliable speaker embedding from polluting the buffer. However, $\mathbf{\hat{y}}_{n}$ can still be adopted because it is predicted by target-speaker embeddings buffered previously. Lastly, valid results of the predicted voice activities will be stitched onto their preceding predictions in the elapsed time. It must be noticed that only the output region belonging to the current chunk is adopted as new predictions in every inference, which ensures the temporal causality of online inference. Valid results of the extracted speaker embeddings are updated in the speaker-embedding buffer to infer the next audio block.

\subsubsection{Buffer Updating}

Fig.~\ref{fig:buffer} illustrates the updating strategies for selecting and buffering target-speaker embeddings at the end of each inference. In this example, both $\hat{w}_{1}$ and $\hat{w}_{2}$ exceed the preset threshold for reserving, but $\hat{w}_{3}$ is discarded. In the dictionary-based speaker-embedding buffer, the keys represent the enrolled speaker labels, and the corresponding values contain lists of embedding-weight pairs, respectively. Each reserved speaker embedding and its weight are appended into the buffer according to the key of the speaker label. When inferring the next audio block, each speaker's target embedding for model input will be the weighted average of all the buffered results. To formally describe this procedure, let $\left \{ \mathbf{\hat{e}}_{n}^{1}, \mathbf{\hat{e}}_{n}^{2}, \dots ,\mathbf{\hat{e}}_{n}^{K_{n}} \right \}$ and $\left \{\hat{w}_{n}^{1}, \hat{w}_{n}^{2}, \dots, \hat{w}_{n}^{K_{n}} \right \}$ denote the embeddings and weights of the $n$-th speaker in the buffer, where $K_{n}$ is the number of embeddings. The aggregation of target-speaker embedding is calculated as follows:
\begin{align}
    \mathbf{\bar{e}}_{n} = \frac{\sum_{k=1}^{K_{n}} (\hat{w}_{n}^{k} \cdot \mathbf{\hat{e}}_{n}^{k})} {\sum_{k=1}^{K_{n}} \hat{w}_{n}^{k}} .
\end{align}

Algorithm~\ref{alg:code} summarizes the pseudocode of online inference in a Python-like style. A live audio signal is fed into the proposed model by a sliding window approach. The neural network detects if a new speaker appears in each coming audio block by itself, eliminating the use of any prior system (e.g., the cascaded diarization). Then, it finishes the target-speaker voice activity detection and embedding extraction for the following audio blocks. In such blockwise processing, the predictions are output immediately as an online diarization system. 

In addition, our proposed framework can achieve better performance through a rescoring mechanism. After the online inference, the speaker-embedding buffer will collect all target-speaker embeddings from the full audio recording. If intermediate features of the extractor and encoder are cached during the first-pass inference, the final speaker-embedding buffer can be used to fastly re-decode the audio, which acts as an offline diarization system. Beneficial to the co-designed training and inferring techniques, our proposed framework adapts to both online and offline inference modes.

\section{Experimental Settings}

\subsection{Datasets}

To train the S2SND models with numerous speaker identities, we introduce two speaker corpora for data simulation. The first corpus is the widely-used VoxCeleb2~\cite{Chung18b} with over 1 million utterances for 6,112 identities. The second corpus is the recently released VoxBlink2~\cite{lin2024voxblink2} with approximately 10 million utterances for 111,284 identities. We employ the FSMN-VAD module in FunASR~\cite{gao2023funasr} toolkit to remove non-speech regions from the raw audio, purifying the data as much as possible. Then, the simulated data is generated in an on-the-fly manner during training. First, the single-speaker utterance is independently created by alternately concatenating the source speech and silent (zero-padded) segments, where each segment length is randomly sampled from a uniform distribution of 0-4 seconds. Second, we randomly mix utterances of 1-3 speakers from the corpora, which follows the same implementation in our previous works~\cite{cheng2023target, cheng2024multi}.

The models pretrained by simulated data are further adapted and evaluated on real multi-domain datasets: DIHARD-II~\cite{ryant2019second} and DIHARD-III~\cite{ryant2020third}, respectively. The DIHARD-II dataset includes 11 conversational scenarios (e.g., interview, clinical, restaurant), with 23.81 hours of development set and 22.49 hours of evaluation set. We select the first 153 recordings (80\%) of the original development set for model adaptation, namely the \textit{dev153} set. The last 39 recordings (20\%) remain for validation, namely the \textit{dev39} set. The DIHARD-III dataset is the next edition of the DIHARD-II dataset in a series of speaker diarization challenges, with 34.15 hours of development set and 33.01 hours of evaluation set. Similarly, we select the first 203 recordings (80\%) of the original development set for model adaptation, namely the \textit{dev203} set. The last 51 recordings (20\%) remain for validation, namely the \textit{dev51} set. The statistics of both simulated and real datasets are described in Table~\ref{tab:data}.

\subsection{Network Configurations}

\subsubsection{Pretrained extractor}

As the pretrained front-end extractor can effectively facilitate the model to learn the identity information in target-speaker embeddings, we pretrain three speaker embedding extractors with similar network architecture but different model sizes and training data. The first two extractors are both based on the ResNet-34 model, while their residual blocks have respective channels of $\left \{32,64,128,256 \right \}$ and $\left \{64,128,256,512 \right \}$, namely the ResNet34-32ch and ResNet34-64ch. After adding the global statistical pooling (GSP)~\cite{snyder2018x} and linear projection layer with the output dimension of 256, these two extractors are trained on the VoxCeleb2~\cite{Chung18b} dataset by the ArcFace ($\alpha=32, m=0.2$)~\cite{deng2019arcface} classifier. We also introduce the third ResNet-152 model trained on the VoxBlink2~\cite{lin2024voxblink2} dataset to explore the potential of large model size and training data. The ResNet34-32ch, ResNet34-64ch, and ResNet-152 models have 5.45M, 21.53M, and 58.14M parameters, respectively. Accordingly, they obtain 1.17\%, 0.81\%, and 0.34\% equal error rates (EERs) on the Vox-O~\cite{Nagrani17} trial.

\subsubsection{S2SND model}

For the entire S2SND model, we propose two versions with different numbers of parameters. The first is named S2SND-Small. Its extractor is based on the ResNet34-32ch model. The following encoder and decoder adopt 256-dim attentions with 8 heads and 512-dim feedforward layers. The second is named S2SND-Medium. Its extractor is based on the ResNet34-64ch model. The encoder and decoder are changed to 384-dim attentions with 8 heads and 768-dim feedforward layers. The other configurations for the two models are identical. All encoders and decoders have 4 blocks. The kernel size of convolutions in Conformer blocks is set to 15. In total, the parameters in the S2SND-Small and S2SND-Medium models are 16.56M and 45.96M, respectively. Because the number of parameters of the ResNet-152 extractor is too large, even twice that of the ResNet34-64ch extractor, the heavy parameters will take too much time to do the experiments and bring high demand for computing cost during real-time inference. We only use the ResNet-152 extractor as the teacher model of knowledge distillation~\cite{hinton2015distilling} to improve the current models described in the following paragraph.

 \begin{table}[t]
	\centering
	\setlength{\tabcolsep}{5pt}
	\caption{Statistics of datasets used in our experiments. The overlap ratios of simulated data are estimated on 250,000 randomly generated samples.}
	%\vspace{-0.2cm}
	\label{tab:data}
	\begin{tabular}{llllr}
		\toprule
		\textbf{Dataset} & \textbf{Split} & \textbf{\makecell[l]{Num. \\ Speakers}} &  \textbf{\makecell[l]{Num. \\ Recordings}} & \textbf{\makecell[r]{Overlap \\ Ratio}}\\
		\midrule
		 \multirow{4}{*}{\makecell{On-the-fly Simulation}} 
		 & sim1spk & 1 & - & 0.00\%  \\
		 & sim2spk & 2 & - & 28.01\% \\
		 & sim3spk & 3 & - & 39.66\% \\
		 \cmidrule(lr){2-5} 
		  & total &1-3 & - & 22.56\% \\
		\midrule
		\multirow{3}{*}{DIHARD-II~\cite{ryant2019second}}
		& dev153 & 1-10  & 153 & 9.78\% \\ 
		& dev39  & 1-9   & 39  & 9.73\% \\
		& eval   & 1-9   & 194 & 8.90\% \\
		\midrule
		DIHARD-III~\cite{ryant2020third} 
		& dev203 & 1-10 & 203 & 10.83\% \\
		& dev51  & 1-8  & 51  & 10.37\% \\
		& eval   & 1-9  & 259 & 9.37\% \\
		\bottomrule
	\end{tabular}
	%\vspace{-0.5cm}
\end{table}

\subsection{Training and Inferring Details}
 
\subsubsection{Training details}

All training audio is split into fixed-length blocks and normalized with a mean of 0 and a standard deviation of 1. Specifically, the block length in this work is set to 8 seconds. The input acoustic features are 80-dim log Mel-filterbank energies with a frame length of 25 ms and a shift of 10 ms. Also, we apply the additive noise from Musan~\cite{snyder2015musan} and reverberation from RIRs~\cite{ko2017study} as audio augmentation. As suggested by our previous findings~\cite{cheng2023target,cheng2024multi}, the temporal resolution (duration per frame-level prediction) of system output is directly set to 10 ms for precise option. The speaker capacity $N$ is adopted as 30, a relatively large number that can adequately cover the maximum number of speakers in most datasets.

When the number of speakers in a given audio block cannot reach $N$, absent positions will be padded as described in Sec.~\ref{sec:masked_speaker_prediction}. Lastly, all the input target-speaker embeddings are randomly shuffled to make the model invariant to speaker order. Accordingly, the ground truth labels for target-speaker voice activity detection and embedding extraction must also be re-assigned based on their shuffled results. Then, the whole model is optimized by AdamW~\cite{loshchilov2017decoupled} optimizer with the binary cross entropy (BCE) loss and ArcFace ($\alpha=32, m=0.2$)~\cite{deng2019arcface} loss depicted in Fig.~\ref{fig:framework}. Using 8 $\times$ NVIDIA RTX-3090 GPUs with a batch size of 16, we investigate two multi-stage training strategies as follows.

The first training strategy follows our previous work~\cite{cheng2023target}, containing three different stages starting from the pretrained extractor. In each stage, the model will be validated every 500 steps. The checkpoint with the lowest diarization error rate on the adopted validation set will be used for the next stage.

\begin{itemize}
\setlength{\itemsep}{0pt}
	\item Stage 1: We copy and freeze the weights of a pretrained speaker embedding model to initialize the front-end extractor. Only simulated data is used to train the back-end modules for 100,000 steps with a learning rate of \textit{1e-4}. 
	\item Stage 2: The front-end extractor is unfrozen. The whole S2SND model is adapted by 80\% of the simulated data and 20\% of the real data from the specific dataset, taking around 75,000 steps.
	\item Stage 3: The learning rate is decayed to \textit{1e-5} for finetuning the whole S2SND model, taking around 50,000 steps.
\end{itemize}

\begin{figure}[t]
\centering
  \includegraphics[width=0.8\linewidth]{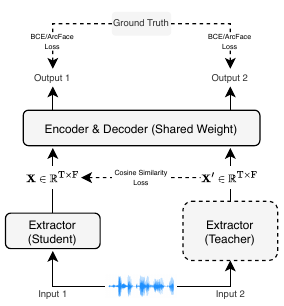}
  \caption{Illustration of the training strategy based on knowledge distillation.}
  \label{fig:distill}
\end{figure}

In this work, we also explore the second kind of training strategy based on knowledge distillation, shown in Fig.~\ref{fig:distill}. The pretrained ResNet-152 model is employed as the teacher extractor. The original input audio will be copied to feed the student and teacher extractors during training. Let $\mathbf{X} = \left [\mathbf{x}_{1}, \ldots , \mathbf{x}_{T} \right ] \in \mathbb{R}^\mathrm{T\times F}$ denote the output of the student extractor, where $T$ is the time axis and $F$ is the feature axis. Comparatively, the output of the teacher extractor is represented as $\mathbf{X}^{\prime} = \left [\mathbf{x}_{1}^{\prime}, \ldots , \mathbf{x}_{T}^{\prime} \right ] \in \mathbb{R}^\mathrm{T\times F}$. Then, we employ a frame-wise cosine similarity loss between two extractors, which is described as: 
\begin{align}
\mathcal{L}_{\mathrm{distill}} = 1 - \frac{1}{T}  \sum_{t=1}^{T} \frac{\mathbf{x}_{t} \cdot \mathbf{x}^{\prime}_{t}} { \| \mathbf{x}_{t} \| \cdot \| \mathbf{x}^{\prime}_{t} \|},
\label{eq:distill} 
\end{align} 
 \noindent where $\mathbf{x}_{t}$ and $\mathbf{x}^{\prime}_{t} \in \mathbb{R}^\mathrm{F}$ represent the frame-level speaker embedding extracted by the student and teacher extractors at time $t$, respectively. By minimizing $\mathcal{L}_{\mathrm{distill}}$, the representation space of $\mathbf{X}$ is forced to align with $\mathbf{X}^{\prime}$, which means the knowledge in the larger teacher extractor transfers into the smaller student extractor. Later, $\mathbf{X}$ and $\mathbf{X}^{\prime}$ are fed into the shared encoder and decoder modules as same as the regular training framework described in Sec.~\ref{sec:training}. The original ground-truth labels are also copied to supervise the two output branches. 
 
During distillation, the total training loss is the sum of $\mathcal{L}_{\mathrm{bce}}$ in Eq.~\ref{eq:bce}, $\mathcal{L}_{\mathrm{arc}}$ in Eq.~\ref{eq:arcface}, and $\mathcal{L}_{\mathrm{distill}}$ in Eq.~\ref{eq:distill}. There are also three training stages, similar to the pretraining strategy.
\begin{itemize}
\setlength{\itemsep}{0pt}
	\item Stage 1: We initialize the weights of the student extractor from scratch and freeze the pretrained teacher extractor. Only simulated data is used to train the student extractor and shared encoder-decoder modules for 100,000 steps with a learning rate of \textit{1e-4}.
	\item Stage 2: The teacher extractor is unfrozen. All weights in Fig.~\ref{fig:distill}, including the student extractor, teacher extractor, and shared encoder-decoder modules, are adapted by 80\% of the simulated data and 20\% of the real data from the specific dataset, taking around 75,000 steps.
	\item Stage 3: The learning rate is decayed to \textit{1e-5} for finetuning based on Stage 2, taking around 50,000 steps.
\end{itemize}

\subsubsection{Inferring details}
\label{sec:inferringr_details}

The inferring process follows the Sec.~\ref{sec:inferring}. The thresholds $\tau_{1}$ and $\tau_{2}$ are determined using grid search on the validation set of the specific dataset. By adjusting the proportion of the current chunk and its contexts in the input audio block, the online diarization system can be flexibly inferred at different latencies. The algorithmic latency is the sum of chunk length $L_\mathrm{chunk}$ and right-context length $L_\mathrm{right}$. As the shift of the sliding window is equal to the chunk length, a smaller $L_\mathrm{chunk}$ can decrease the system latency but bring intensive computing. The right context represents the use of future information. A larger $L_\mathrm{right}$ may result in more accurate prediction but increase the system latency. The impacts of different settings are investigated in the experimental results.

\subsection{Evaluation Metric}

The diarization error rate (DER) is used as the evaluation metric without collar tolerance. The S2SND models are tested on evaluation sets of DIHARD-II~\cite{ryant2019second} and DIHARD-III~\cite{ryant2020third} datasets. For a fair comparison, the Oracle VAD information can revise the diarization results as a post-processing approach~\cite{horiguchi2022encoder} if sometimes the evaluation condition allows.

\begin{table*}[t]
	\centering
	\caption{Performance of S2SND models on DIHARD-II and DIHARD-III evaluation sets with various training and inferring conditions. The Diarization Error Rates (DERs) are reported without Oracle VAD and collar tolerance.}
	%\vspace{-0.2cm}
	\label{exp:dihard}
	\begin{threeparttable}[b]
	\begin{tabular}{lllllllrrrr}
		\toprule
		\multirow{3}{*}{\textbf{ID}}
		& \multirow{3}{*}{\textbf{\makecell[l]{Model\\Size}}}
		& \multirow{3}{*}{\textbf{\makecell[l]{Simulation\\Corpus}}}
		& \multirow{3}{*}{\textbf{\makecell[l]{Training\\Strategy}}}
		& \multirow{3}{*}{\textbf{\makecell[l]{Chunk\\Length}}}
		& \multirow{3}{*}{\textbf{\makecell[l]{Right-Context\\Length}}}
		& \multirow{3}{*}{\textbf{\makecell[l]{Algorithmic\\Latency}}}
		& \multicolumn{2}{c}{\textbf{DIHARD-II Eval}}
		& \multicolumn{2}{c}{\textbf{DIHARD-III Eval}}
		\\
		\cmidrule(lr){8-9}
		\cmidrule(lr){10-11}
		&&&&&&& \textbf{\makecell[l]{Online\\DER (\%)}} & \textbf{\makecell[l]{Offline\\DER (\%)}} & \textbf{\makecell[l]{Online\\DER (\%)}} & \textbf{\makecell[l]{Offline\\DER (\%)}}
		\\
        \midrule
		S1
	    & \multirow{4}{*}{\makecell[l]{Small}}
	    & \multirow{4}{*}{VoxCeleb2}
	    & \multirow{4}{*}{Pretraining} 
	    &    0.48s & -     & 0.48s & 27.79 & 23.74 & 20.55 & \cellcolor{gray!25} 16.28 \\
	    S2 &&&& 0.48s & 0.16s & 0.64s & 26.11 & 23.85 & 18.53 & 16.33 \\
	    S3 &&&& 0.64s & -     & 0.64s & 27.54 & 24.07 & 19.72 & 16.36 \\
	    S4 &&&& 0.64s & 0.16s & 0.80s & 25.79 & 23.52 & \cellcolor{gray!25} 18.33 & 16.33 \\
	    
	    \midrule
		S5
	    & \multirow{4}{*}{\makecell[l]{Small}}
	    & \multirow{4}{*}{VoxBlink2}
	    & \multirow{4}{*}{Pretraining} 
	    &    0.48s & -     & 0.48s & 27.39 & \cellcolor{gray!25} 22.80 & 20.45 & 16.51 \\
	    S6 &&&& 0.48s & 0.16s & 0.64s & 25.72 & 22.97 & 18.44 & 16.38 \\
	    S7 &&&& 0.64s & -     & 0.64s & 26.93 & 23.05 & 19.55 & 16.32 \\
	    S8 &&&& 0.64s & 0.16s & 0.80s & \cellcolor{gray!25} 25.70 & 23.17 & 18.36 & 16.45 \\
	  
	    \midrule
		S9 
	    & \multirow{4}{*}{\makecell[l]{Small}}
	    & \multirow{4}{*}{VoxBlink2}
	    & \multirow{4}{*}{Distillation} 
	    &    0.48s & -     & 0.48s & 27.87 & 23.88 & 21.33 & 17.70 \\
	    S10 &&&& 0.48s & 0.16s & 0.64s & 26.71 & 24.36 & 19.42 & 17.51 \\
	    S11 &&&& 0.64s & -     & 0.64s & 27.58 & 24.28 & 20.67 & 17.57 \\
	    S12 &&&& 0.64s & 0.16s & 0.80s & 26.21 & 24.29 & 19.23 & 17.57 \\   

        \midrule
        \midrule
		S13 
	    & \multirow{4}{*}{\makecell[l]{Medium}}
	    & \multirow{4}{*}{VoxCeleb2}
	    & \multirow{4}{*}{Pretraining} 
	    &    0.48s & -     & 0.48s & 27.57 & 23.78 & 20.61 & 16.81 \\
	    S14 &&&& 0.48s & 0.16s & 0.64s & 25.57 & 23.61 & 18.82 & 16.97 \\
	    S15 &&&& 0.64s & -     & 0.64s & 27.00 & 23.83 & 20.08 & 16.83 \\
	    S16 &&&& 0.64s & 0.16s & 0.80s & 25.78 & 23.89 & 18.43 & 16.79 \\
  	   
	    \midrule
		S17 
	    & \multirow{4}{*}{\makecell[l]{Medium}}
	    & \multirow{4}{*}{VoxBlink2}
	    & \multirow{4}{*}{Pretraining} 
	    &    0.48s & -     & 0.48s & 27.79 & 24.09 & 20.13 & 16.04 \\
	    S18 &&&& 0.48s & 0.16s & 0.64s & 26.02 & 23.86 & 18.10 & 15.77 \\
	    S19 &&&& 0.64s & -     & 0.64s & 27.10 & 23.77 & 19.30 & 15.83 \\
	    S20 &&&& 0.64s & 0.16s & 0.80s & 25.55 & 23.59 & 17.99 & 15.93 \\

	    \midrule
		S21 
	    & \multirow{4}{*}{\makecell[l]{Medium}}
	    & \multirow{4}{*}{VoxBlink2}
	    & \multirow{4}{*}{Distillation} 
	    &    0.48s & -     & 0.48s & 26.13 & \cellcolor{gray!25} 21.95 & 19.11 & 15.14 \\
	    S22 &&&& 0.48s & 0.16s & 0.64s & \cellcolor{gray!25} 24.41 & 22.17 & 17.33 & 15.30 \\
	    S23 &&&& 0.64s & -     & 0.64s & 25.44 & 22.07 & 18.46 & \cellcolor{gray!25} 15.13 \\
	    S24 &&&&  0.64s & 0.16s & 0.80s & 24.48 & 22.26 & \cellcolor{gray!25} 17.12 & 15.23 \\       	   	    
		\bottomrule
	\end{tabular}
	\begin{tablenotes}
		\item The lowest online and offline DERs of each model size are highlighted by the gray background. 
    \end{tablenotes}
	\end{threeparttable}
	%\vspace{-0.5cm}
\end{table*}

\section{Results}
\label{sec:results}

\subsection{Evaluation of S2SND Models}

Table~\ref{exp:dihard} illustrates the performance of our proposed S2SND models with different training and inferring conditions. The effects of model size, simulation corpus, training strategy, and various combinations of chunk and right-context lengths are shown step by step. Browsing the DER results on DIHARD-II and DIHARD-III datasets, several consequences are found as follows.

\begin{enumerate}
\setlength{\itemsep}{0pt}
	\item Across all experimental groups, under the same conditions (e.g., model size, simulation corpus, and training strategy), the longer chunk and right-context lengths can generally result in lower online DERs. Especially, using right-context information means that future information is exploited when predicting each chunk, which leads to a more significant impact. On the other hand, although the chunk length has less influence on the DERs, adjusting it can help maintain the algorithmic latency constant while increasing the right-context length. For instance, \textit{S2} has lower online DERs than \textit{S3}, even though their total latencies are equal. These phenomena are also shown in all the other experimental groups. To avoid too large system latency during online inference, the chunk and right-context lengths used in our experiments are selected to be relatively small and close values. Furthermore, the longer chunk and right-context lengths do not exhibit apparent advantages in offline DERs. The rescoring mechanism updates the diarization output over the whole recording, which already takes global information. Its offline performance is insensitive to the context length of the first-pass online inference.
	\item Comparing \textit{S1-4}, \textit{S5-8}, and \textit{S9-12}, we evaluate the small model with different simulation corpora and training strategies. It can be seen that the VoxBlink2 corpus containing larger speaker identities (111k+) does not result in significant and consistent improvement over VoxCeleb2 (6k+). Also, the training strategy of knowledge distillation slightly downgrades the performance compared to the pretraining strategy. It is speculated that the small model with few parameters cannot fully exploit the large simulation corpus and knowledge distillation.
	\item Comparing \textit{S13-16}, \textit{S17-20} and \textit{S21-24}, we evaluate the medium model with different simulation corpora and training strategies again. In this case, the combination of VoxBlink2 corpus and knowledge distillation demonstrates overwhelming advantages over others. All the lowest DERs for medium model on two datasets are obtained in \textit{S21-24}. When increasing the number of model parameters, the newly introduced distillation strategy can successfully empower the usage of large speaker identities in the simulation corpus.
\end{enumerate}

\begin{table}[t]
	\centering
	\setlength{\tabcolsep}{1.5pt}
	\caption{Comparisons of S2SND models with others on the DIHARD-II evaluation set.}
	%\vspace{-0.2cm}
	\label{dihard2_comp}
	\begin{threeparttable}[b]
	\begin{tabular}{lrr}
		\toprule
		\textbf{Method} & \textbf{Latency (s)} & \textbf{DER (\%)} \\	    
		\midrule
		\textbf{Online} \\
		\quad EEND-EDA + FW-STB~\cite{xue2021online_2} & 1.00 & 36.00 \\
		\quad EEND-EDA + Improved FW-STB~\cite{horiguchi2022online} & 1.00 & 33.37 \\	
		\quad Overlap-aware Speaker Embeddings ~\cite{coria2021overlap} & 1.00 & 35.10 \\
		\quad EEND-GLA-Small + BW-STB~\cite{horiguchi2022online} & 1.00 & 31.47 \\
		\quad EEND-GLA-Large + BW-STB~\cite{horiguchi2022online} & 1.00 & 30.24 \\
	    % ---------------- % 
	    \quad S2SND-Small (S8 in Table~\ref{exp:dihard}) & 0.80 &  25.70 \\	
        \quad S2SND-Medium (S22 in Table~\ref{exp:dihard}) & 0.64 &  \textbf{24.41} \\	
	    % ---------------- % 
        \textbf{Online (with oracle voice activity detection)} \\
	 	\quad UIS-RNN-SML~\cite{fini2020supervised} & 1.00 & 27.30 \\
	 	\quad EEND-EDA + FW-STB~\cite{xue2021online_2} & 1.00 & 25.80 \\
	 	\quad EEND-EDA + Improved FW-STB~\cite{horiguchi2022online} & 1.00 & 24.67 \\
	 	\quad Core Samples Selection~\cite{yue2022online} & 1.00 & 23.10 \\
	 	\quad EEND-GLA-Small + BW-STB~\cite{horiguchi2022online} & 1.00 & 23.26 \\
	 	\quad EEND-GLA-Large + BW-STB~\cite{horiguchi2022online} & 1.00 & 21.92 \\
	 	\quad NAVER System~\cite{kwon2023absolute} & 0.50 & 21.60\\
	    % ---------------- % 
	    \quad S2SND-Small (S8 in Table~\ref{exp:dihard}) + Oracle VAD & 0.80 & \textbf{18.07} \\	
        \quad S2SND-Medium (S22 in Table~\ref{exp:dihard}) + Oracle VAD & 0.64 & 18.65 \\	
	    % ---------------- % 
	    
		\midrule	
		\textbf{Offline} \\
		\quad EEND-EDA~\cite{horiguchi2022encoder} &  & 29.57 \\	
	 	\qquad + Iterative Inference+~\cite{horiguchi2022encoder} &  & 28.52 \\
		\quad EEND-GLA-Small~\cite{horiguchi2022online} &  & 29.31 \\
	 	\quad EEND-GLA-Large~\cite{horiguchi2022online} &  & 28.33 \\
		\quad BUT System~\cite{landini2020but}~\tnote{$\dag$} &  & 27.11 \\
	 	\qquad + EEND Post-Processing~\cite{horiguchi2021end} &  & 26.88 \\
	 	\quad AED-EEND~\cite{chen2024attention} &  & 25.92 \\
	 	\qquad + Embedding Enhancer~\cite{chen2024attention} &  & 24.64 \\
	 	% ---------------- % 
	 	\quad S2SND-Small (S5 in Table~\ref{exp:dihard}) &  & 22.80 \\
	 	\quad S2SND-Medium (S21 in Table~\ref{exp:dihard}) &  & \textbf{21.95} \\
		% ---------------- % 
	    \textbf{Offline (with oracle voice activity detection)}  \\
	    \quad EEND-EDA~\cite{horiguchi2022encoder} &  & 20.54 \\
	    \qquad + Iterative Inference+~\cite{horiguchi2022encoder} &  & 20.24 \\
	    \quad VBx~\cite{landini2022bayesian} & & 18.55 \\
	    \quad BUT System~\cite{landini2020but}~\tnote{$\dag$} & & 18.42 \\
	    % ---------------- % 
	  	\quad S2SND-Small (S5 in Table~\ref{exp:dihard}) + Oracle VAD &  & 15.84 \\
	 	\quad S2SND-Medium (S21 in Table~\ref{exp:dihard}) + Oracle VAD &  & \textbf{15.34} \\    
	    % ---------------- % 
	    
		\bottomrule
	\end{tabular}
	\begin{tablenotes}
		\item[$\dag$] Winning system on Track 1\&2 of the DIHARD-II Challenge. 
    \end{tablenotes}
	\end{threeparttable}
	%\vspace{-0.5cm}
\end{table}

\begin{table}[t]
	\centering
	\setlength{\tabcolsep}{1.5pt}
	\caption{Comparisons of S2SND models with others on the DIHARD-III evaluation set.}
	%\vspace{-0.2cm}
	\label{dihard3_comp}
	\begin{threeparttable}[b]
	\begin{tabular}{lrr}
		\toprule
		\textbf{Method} & \textbf{Latency (s)} & \textbf{DER (\%)} \\
        \midrule
		\textbf{Online} \\
        \quad Overlap-aware Speaker Embeddings ~\cite{coria2021overlap} & 1.00 & 27.60 \\
        \quad EEND-EDA + Improved FW-STB~\cite{horiguchi2022online} & 1.00 & 25.09 \\
        \quad EEND-GLA-Small + BW-STB~\cite{horiguchi2022online} & 1.00 & 22.00 \\
		\quad EEND-GLA-Large + BW-STB~\cite{horiguchi2022online} & 1.00 & 20.73 \\
		\quad ResNet-based OTS-VAD~\cite{wang2023end} & 0.80 & 19.07 \\
		% ---------------- %
		\quad S2SND-Small (S4 in Table~\ref{exp:dihard}) & 0.80 & 18.33 \\
		\quad S2SND-Medium (S24 in Table~\ref{exp:dihard}) & 0.80 & \textbf{17.12} \\
		% ---------------- %
        \textbf{Online (with oracle voice activity detection)} \\
        \quad Zhang et al.~\cite{zhang22_odyssey} & 0.50 & 19.57 \\
        \quad Core Samples Selection~\cite{yue2022online} & 1.00 & 19.30 \\
        \quad NAVER System~\cite{kwon2023absolute} & 0.50 & 19.05 \\
        \quad EEND-EDA + Improved FW-STB~\cite{horiguchi2022online} & 1.00 & 18.58 \\
        \quad EEND-GLA-Small + BW-STB~\cite{horiguchi2022online} & 1.00 & 15.82 \\
		\quad EEND-GLA-Large + BW-STB~\cite{horiguchi2022online} & 1.00 & 14.70 \\
        \quad ResNet-based OTS-VAD~\cite{wang2023end} & 0.80 & 13.31 \\
		% ---------------- %
		\quad S2SND-Small (S4 in Table~\ref{exp:dihard}) + Oracle VAD & 0.80 & 13.07 \\
		\quad S2SND-Medium (S24 in Table~\ref{exp:dihard}) + Oracle VAD & 0.80 & \textbf{11.88} \\

		% ---------------- %
		\midrule	
		\textbf{Offline} \\	
		\quad EEND-EDA~\cite{horiguchi2022encoder} &  &  21.55 \\
        \qquad + Iterative Inference+~\cite{horiguchi2022encoder} &  & 20.69 \\
        \quad Pyannote.audio v3.1~\cite{plaquet2023powerset} &  & 21.30 \\
		\quad DiaPer~\cite{landini2024diaper} &  & 20.30 \\
        \quad EEND-GLA-Small~\cite{horiguchi2022online} &  & 20.23 \\
	 	\quad EEND-GLA-Large~\cite{horiguchi2022online} &  & 19.49 \\
		\quad VBx + Overlap-aware Resegmentation~\cite{bredin2021end} &  & 19.30 \\
        \quad USTC-NELSLIP System~\cite{wang2021ustc}~\tnote{$\dag$} &  & 16.78 \\ 
        \quad ANSD-MA-MSE~\cite{he2023ansd} & & 16.76 \\   
        \quad EEND-M2F~\cite{harkonen2024eend} & & 16.07 \\
        % ---------------- % 
        \quad S2SND-Small (S1 in Table~\ref{exp:dihard}) &  & 16.28 \\	
        \quad S2SND-Medium (S23 in Table~\ref{exp:dihard}) &  & \textbf{15.13} \\
        % ---------------- % 
        \textbf{Offline (with oracle voice activity detection)}  \\
        \quad EEND-EDA~\cite{horiguchi2022encoder} &  & 14.91 \\
        \quad + Iterative Inference+~\cite{horiguchi2022encoder} &  & 14.42 \\
        \quad Hitachi-JHU System~\cite{horiguchi2021hitachi} & & 11.58 \\
        \quad USTC-NELSLIP System~\cite{wang2021ustc}~\tnote{$\dag$} & & 11.30 \\
        \quad ANSD-MA-MSE~\cite{he2023ansd} & & 11.12 \\
        \quad Seq2Seq-TSVAD~\cite{cheng2023target} & & 10.77 \\
        \quad MIMO-TSVAD~\cite{cheng2024multi} & & \textbf{10.10} \\ 
         % ---------------- % 
        \quad S2SND-Small (S1 in Table~\ref{exp:dihard}) + Oracle VAD &  & 11.13 \\	
        \quad S2SND-Medium (S23 in Table~\ref{exp:dihard}) + Oracle VAD &  & 10.37 \\
        % ---------------- % 
		\bottomrule
	\end{tabular}
	\begin{tablenotes}
		\item[$\dag$] Winning (fusion) system on Track 1\&2 of the DIHARD-III Challenge.  
    \end{tablenotes}
	\end{threeparttable}
	%\vspace{-0.5cm}
\end{table}

Overall, for the S2SND-Small model, the best online DERs on DIHARD-II and DIHARD-III datasets are 25.70\% and 18.33\%, and the best offline DERs on the two datasets are 22.80\% and 16.28\%, respectively. For the S2SND-Medium model, the best online DERs on DIHARD-II and DIHARD-III datasets are 24.41\% and 17.12\%, and the best offline DERs on two datasets are 21.95\% and 15.13\%, respectively. To summarize, the combination of a small simulation corpus and pretraining strategy is the better choice for the small model. When a large simulation corpus is available, adopting the medium model and distillation strategy can achieve better DER performance.

\subsection{Comparison with Other Existing Methods}

We select the lowest online and offline DERs for each model size on DIHARD-II and DIHARD-III datasets as the representative results, highlighted by the gray background in Table~\ref{exp:dihard}. To fairly compare with some existing methods, the corresponding results of post-processing by Oracle VAD ~\cite{horiguchi2022encoder} are also provided.

Table~\ref{dihard2_comp} compares our proposed methods with the previous state-of-the-art results on the DIHARD-II dataset. In the online scenario, our proposed methods obtain the lowest DERs of 18.07\% and 24.41\% with and without Oracle VAD, respectively. Regarding algorithmic latency, our proposed methods still have a significant advantage over others when Oracle VAD is not used. In the offline scenario, our proposed methods obtain the lowest DERs of 15.34\% and 21.95\% with and without Oracle VAD, respectively. Generally, our best results significantly outperform previous state-of-the-art systems in all scenarios. Notably, our best online DER (24.41\%) is even lower than the previous best offline system (24.64\%)~\cite{chen2024attention} and the winning system (27.11\%)~\cite{landini2020but} of the DIHARD-II Challenge.

Table~\ref{dihard3_comp} compares our proposed methods with the previous state-of-the-art results on the DIHARD-III dataset. In the online scenario, our proposed methods obtain the lowest DERs of 11.88\% and 17.12\% with and without Oracle VAD, respectively. In the offline scenario, our proposed methods obtain the lowest DERs of 10.37\% and 15.13\% with and without Oracle VAD, respectively. Except for the offline result with Oracle VAD, our best results significantly outperform previous state-of-the-art systems in all other scenarios. Nevertheless, the DER (10.10\%) of MIMO-TSVAD~\cite{cheng2024multi} comes from our earlier study designed for offline scenarios, which adopts the audio block of 32 seconds to provide extended context but is not suitable for online inference. Last but not least, our best online DER (17.12\%) is very close to the previous best offline system (16.07\%)~\cite{harkonen2024eend} and the winning fusion system (16.78\%)~\cite{wang2021ustc} of the DIHARD-III Challenge.

\subsection{Investigation of Speaker Counting Ability}

The previous speaker diarization systems mainly utilize unsupervised clustering~\cite{wang2018speaker,lin2019lstm,landini2022bayesian,wang2022similarity}, permutation-invariant training~\cite{fujita2019end_1,fujita2019end_2, horiguchi2020end,horiguchi2022encoder}, or their combination to determine the unknown number of speakers in the input audio. In our proposed S2SND framework, speakers are detected by traversing the entire audio using the masked speaker prediction mechanism, which is clustering-free. Also, it only adds one unknown speaker each time to avoid the increasing complexity problem of the permutation-invariant training.

\begin{figure*}[t]
\centering
	\subfloat[Pyannote.audio v3.1~\cite{plaquet2023powerset}]{\includegraphics[width=0.24\linewidth]{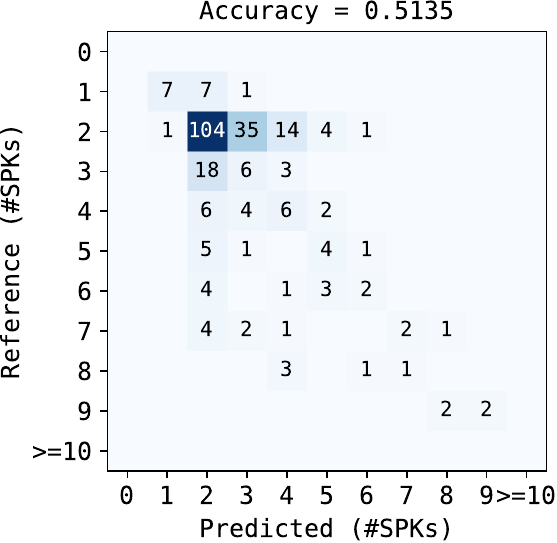}
	\label{pyannote}}
	\subfloat[VBx~\cite{landini2022bayesian}]{\includegraphics[width=0.24\linewidth]{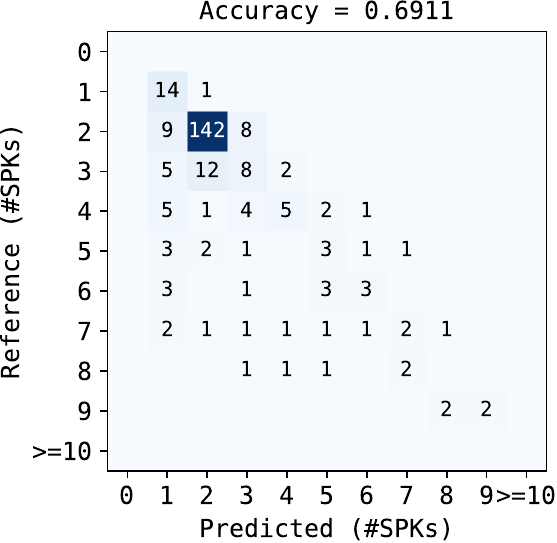}
	\label{vbx}}
	\subfloat[DiaPer~\cite{landini2024diaper}]{\includegraphics[width=0.24\linewidth]{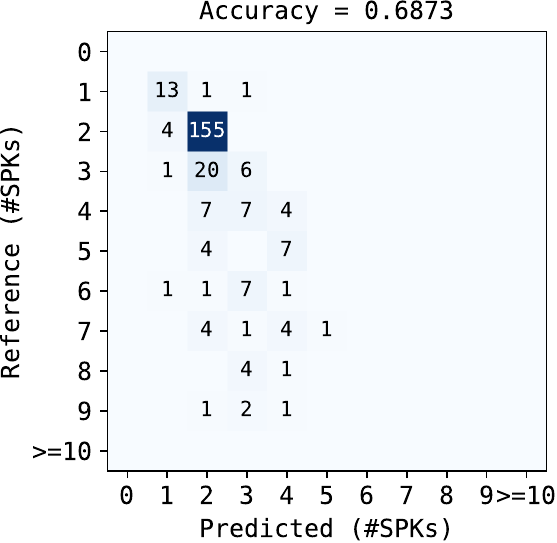}
	\label{diaper}}
	\subfloat[S2SND-Small (Offline)]{\includegraphics[width=0.24\linewidth]{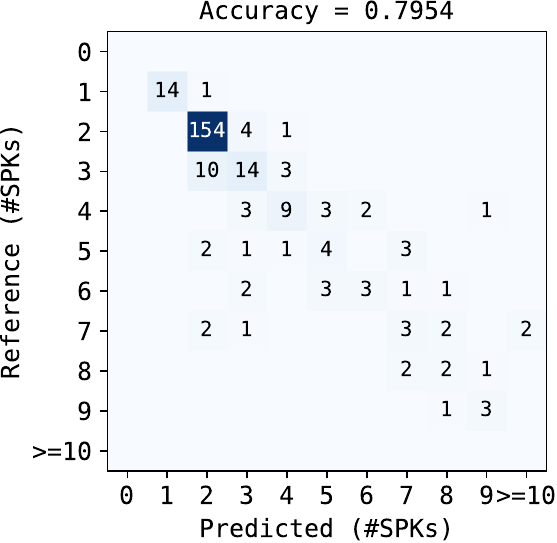}
	\label{s2snd_small_offline}}
\caption{Confusion matrices for speaker counting on the DIHARD-III evaluation set. The Pyannote.audio v3.1, VBx, and DiaPer results are provided by their respective authors. For our trained S2SND-Small model, the same settings as \textit{S4} in Table~\ref{exp:dihard} are adopted. Oracle VAD is not used. Accuracy is calculated as the number of recordings in which all speakers are correctly predicted, divided by the total number of recordings.}
\label{fig_cm}
%\vspace{-0.3cm}
\end{figure*}

Fig.~\ref{fig_cm} depicts the confusion matrices for speaker counting obtained by different methods on the DIHARD-III evaluation set. Due to space limitations, only offline performances without Oracle VAD are shown. We select three representative systems from different technical routes for comparison. The first Pyannote.audio v3.1~\cite{plaquet2023powerset} is a hybrid method of supervised end-to-end diarization and unsupervised clustering. The second VBx~\cite{landini2022bayesian} is a well-known clustering-based diarization method, where the shown performance is reproduced as the baseline in the third EEND-based method (DiaPer~\cite{landini2024diaper}). As a result, the S2SND-Small model exhibits the more balanced predictions with the highest accuracy of 79.54\%, proving the speaker counting ability of our proposed S2SND framework.

\subsection{Computing Efficiency}

Table~\ref{efficiency} illustrates the computing efficiency of the S2SND models. First, the total number of model parameters is an essential measurement. Second, as mentioned in Sec~\ref{sec:inferringr_details}, the chunk length ($L_\mathrm{chunk}$) determines the shift of the sliding window in our settings. Given the amount of floating-point operations for processing each window as $\Delta_\mathrm{flops}$, the Floating-Point Operations Per Second (FLOPS) is calculated as $\Delta_\mathrm{flops}/L_\mathrm{chunk}$. A smaller chunk length (window shift) means more windows must be processed per unit time, leading to more FLOPS because of more intensive computing. On the contrary, a larger chunk length (window shift) is more computationally economical, but the system latency will increase. Third, the Real-time Factor (RTF) is calculated as the time to process each recording divided by the recording length, where all tests are based on the computer with Intel(R) Xeon(R) E5-2660 CPU @ 2.60GHz and NVIDIA RTX-3090 GPU.

When using GPU inference, the maximum RTF of 0.22 is adequate for real-time applications. Also, when using CPU inference, the maximum RTF of 0.60 is not an excellent performance, but it is less than 1, which means the system operation is still in real-time. Our calculations of RTFs include all the time the speaker diarization system runs, not only the neural network inference but also the actual data I/O, signal preprocessing, buffer update, etc. Therefore, the RTFs tested on the CPU are not much larger than that of the GPU, especially for the small model size. From the perspective of computing efficiency, our proposed S2SND models are not outstanding. Nevertheless, they achieve promising diarization performance (as shown in Tables~\ref{dihard2_comp} and~\ref{dihard3_comp}) with improved speaker counting performance (as shown in Fig.~\ref{fig_cm}). In our future work, we will further improve the computational efficiency.

\begin{table}[t]
	\centering
	\caption{Computing Efficiency regarding the number of parameters, Floating-Point Operations Per Second (FLOPS), and Real-Time Factor (RTF).}
	%\vspace{-0.2cm}
	\label{efficiency}
	\begin{threeparttable}[b]
	\begin{tabular}{lrrrrrr}
		\toprule
		\multirow{2}{*}{\textbf{Model}} 
		& \multirow{2}{*}{\textbf{\makecell[r]{Params\\(M)}}}
		& \multirow{2}{*}{\textbf{\makecell[r]{FLOPS\\(G)}}}
		& \multirow{2}{*}{\textbf{\makecell[r]{RTF-\\GPU}}}
		& \multirow{2}{*}{\textbf{\makecell[r]{RTF-\\CPU}}}
		\\
		\\
        \midrule
        S2SND-Small \\
        \quad $L_\mathrm{chunk}=0.48s$  & 16.56 & 78.83  & 0.19 &  0.34 \\
        \quad $L_\mathrm{chunk}=0.64s$  & 16.56 & 59.12  & 0.14 &  0.21 \\
        S2SND-Medium \\
        \quad $L_\mathrm{chunk}=0.48s$ & 45.96 & 308.89  & 0.22 &  0.60 \\
        \quad $L_\mathrm{chunk}=0.64s$ & 45.96 & 231.67  & 0.15 &  0.39 \\   
		\bottomrule
	\end{tabular}
	\end{threeparttable}
	%\vspace{-0.5cm}
\end{table}

Furthermore, comparing computing efficiency depends on various measurement criteria and hardware platforms. For instance, EEND-GLA-Small~\cite{horiguchi2022online} has only 6.4M parameters. However, it additionally relies on clustering of relative speaker embeddings, which has $\mathcal{O}(n^{3})$ time complexity but cannot be counted into FLOPS on the GPU device. OTS-VAD~\cite{wang2023end} employs an external VAD module to remove silent regions from the original audio signal. The preprocessing time (e.g., VAD) is not involved in the RTFs reported by the authors. It is hard to compare different studies in those aspects fairly. Thus, this paper does not list the computing efficiency comparisons with other methods.

\section{Conclusions}
This paper proposes a novel Sequence-to-Sequence Neural Diarization (S2SND) framework to tackle online and offline speaker diarization in a unified model. The S2SND models can automatically detect and represent an unknown number of speakers in the input audio signal using the well-designed training and inferring process. Experimental results show that the proposed S2SND framework obtains new state-of-the-art DERs across all online and offline inference scenarios. Nevertheless, the proposed models also have limitations. The large model size and computing cost still present challenges for real-time inference on edge devices without GPUs. In the future, we will further improve the current approach regarding both precision and speed, prompting speaker diarization to wide industrial applications.

\section*{Acknowledgments}
This research is funded in part by the National Natural Science Foundation of China (62171207), Yangtze River Delta Science and Technology Innovation Community Joint Research Project (2024CSJGG01100), Science and Technology Program of Suzhou City (SYC2022051) and Guangdong Science and Technology Plan (2023A1111120012). Many thanks for the computational resource provided by the Advanced Computing East China Sub-Center.

% References section
%
\bibliographystyle{IEEEtran}
\bibliography{refs}

\end{document}